\documentclass{article}
\usepackage{arxiv}

\usepackage[utf8]{inputenc} 
\usepackage[T1]{fontenc}    
\usepackage{hyperref}       
\usepackage{url}            
\usepackage{booktabs}       
\usepackage{amsfonts}       
\usepackage{nicefrac}       
\usepackage{microtype}      
\usepackage{lipsum}
\usepackage{float}
\usepackage{graphicx}
\usepackage{multirow}
\usepackage{array}
\usepackage{comment}
\usepackage{longtable}
\usepackage{ragged2e}
\usepackage{subfig}

\title{Industry 4.0 in Health care: A systematic review}

\author{
  Md Manjurul Ahsan \\
  Industrial and Systems Engineering\\
  University of Oklahoma\\
  Norman, Oklahoma-73071 \\
  \texttt{ahsan@ou.edu} \\
   \And
 Zahed Siddique \\
  Department of Aerospace and Mechanical Engineering\\
  University of Oklahoma\\
  Norman, Oklahoma-73071\\
  \texttt{zsiddique@ou.edu}} 

\begin{document}
\maketitle

\begin{abstract}
Industry 4.0 in health care has evolved drastically over the past century. In fact, it is evolving every day, with new tools and strategies being developed by physicians and researchers alike. Health care and technology have been intertwined together with the advancement of cloud computing and big data. This study aims to analyze the impact of industry 4.0 in health care systems. To do so, a systematic literature review was carried out considering peer-reviewed articles extracted from the two popular databases: Scopus and Web of Science (WoS). PRISMA statement 2015 was used to include and exclude that data. At first, a bibliometric analysis was carried out using 346 articles considering the following factors: publication by year, journal, authors, countries, institutions, authors' keywords, and citations. Finally, qualitative analysis was carried out based on selected 32 articles considering the following factors: a conceptual framework, schedule problems, security, COVID-19, digital supply chain, and blockchain technology. Study finding suggests that during the onset of COVID-19, health care and industry 4.0 has been merged and evolved jointly, considering various crisis such as data security, resource allocation, and data transparency. Industry 4.0 enables many technologies such as the internet of things (IoT), blockchain, big data, cloud computing, machine learning, deep learning, information, and communication technologies (ICT) to track patients' records and helps reduce social transmission COVID-19 and so on. The study findings will give future researchers and practitioners some insights regarding the integration of health care and Industry 4.0.
\end{abstract}

\section{Introduction}\label{sec1}
Industry 4.0, is the fourth industrial revolution~\cite{gorecki2021business} since the introduction of electricity to factories in the 19$^{th}$ century and Internet technology to business in the 90s.
The digitization of business processes and the Internet of Things (IoT) have already started transforming the world of manufacturing, logistics, and supply chain management~\cite{manavalan2019review}; however, health care has been lagging behind in terms of digitalization and innovation~\cite{golinelli2020adoption}. Thanks to the proliferation of mobile devices and medical instruments, healthcare technologies can now connect and exchange data with each other on an unprecedented scale~\cite{chen2012challenges}. The health care industry has made several major transformations over the past few decades, from introducing vaccines and antibiotics to creating patient-centered care and now to next-generation genomic medicine~\cite{american2014innovation}. At each stage in this evolution, we have seen an explosion of innovation that changed how patients and providers interact with one another.\\
As the establishment of tools grows, health care systems experience radical modifications and technical specifications concerning industry 4.0~\cite{kumar2019industry}. Industrial revolution 4.0 or, in short, IR 4.0 emerges significantly during the onset of COVID-19. Factors such as social distancing and lockdown in various nations significantly affected almost every country's economy and industrial domain~\cite{anjum2020mapping,ahsan2021detection}. As a result, a compulsory merger of IR 4.0 was observed more during the onset of COVID-19 years compared to previous decades. The deadly COVID-19 brings forth many healthcare lackings worldwide, and many countries like Bangladesh, India, the USA, and the UK failed to facilitate health care systems appropriately~\cite{ahsan2021detecting,ahsan2020deep,ahsan2020covid}. \\
However, technological advancement plays a vital role during this pandemic. In order to keep up with the demands of their patients, many healthcare professionals are turning to new technologies, including mobile apps and remote monitoring, to improve the quality of patient care they provide. For example, telemedicine allows nurses and physicians to reach patients in rural or isolated areas who might otherwise have trouble finding adequate medical care, while wearable technology helps to monitor cardiac conditions~\cite{ahsan2021effect} and other ailments over time. These new tools allow for more efficient services that can provide medical services for distant residents and reduce the loads on the hospitals and emergency rooms~\cite{mustapha2021impact}.\\
The advent of COVID-19 and the adoption of IR 4.0 in health care generates various directions. Therefore, there is a need to integrate all those directions in a nutshell. Considering this opportunity into account, this study aims to conduct a systematic review of recently published literature that concentrated on two key factors: health care and Industry 4.0.\\
The literature on Industry 4.0 in health care (IHC) is expanding at an exponential rate (see Fig.~\ref{fig:fig3}), highlighting the need to transmit new insights and directions based on the most current advances in the existing research. Thus, 346 publications were analyzed bibliometrically to answer the following study questions: who are the notable authors? What is the current trend in publishing? What are the most influential journals? Which countries, institutions, and topics make the most contributions? Furthermore, from which essential concepts may the extant IHC literature be classified?\\
Furthermore, in order to provide some insight into IHC, the study intended to answer the following questions using 32 referenced literature:
\begin{enumerate}
    \item What are the most important applications of Industry 4.0 in health care?
    \item What are the current Industry 4.0 challenges and constraints in health care?
    \item What are the potential future applications of Industry 4.0 in health care?
\end{enumerate}

It should be noted that all 32 references were retrieved using the inclusion criteria as outlined in Section~\ref{method}.The remainder of the section is organized as follows: Section~\ref{analysis} describes the bibliometric study based on the relevant literature, and Section~\ref{HCI} presents the IHC findings. Finally, Section~\ref{con} summarizes the general findings and offers some suggestions for future study and practice areas for researchers and practitioners.
\section{Methodology}\label{method}
\begin{figure}[htbp]
    \centering
    \includegraphics[width=\textwidth]{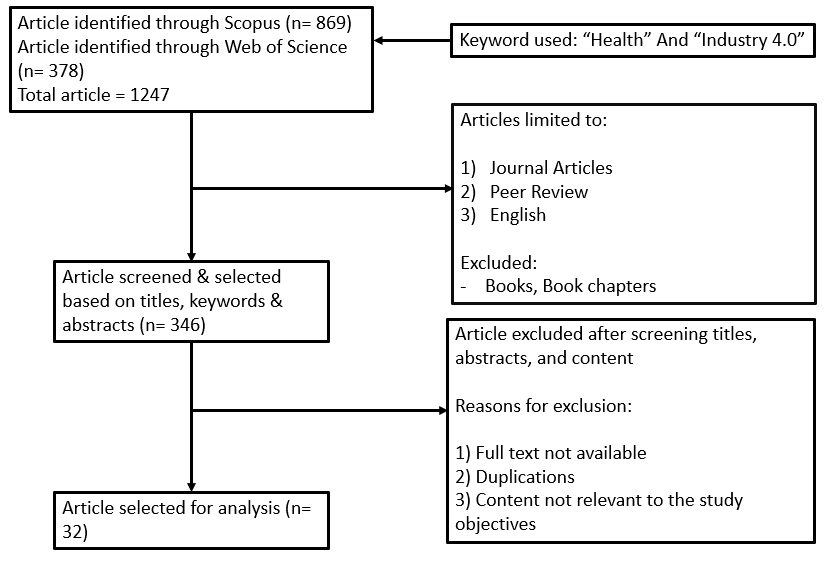}
    \caption{Flow diagram of article selection procedure used in this study using PRISMA 2015~\cite{moher2015preferred}}
    \label{fig:fig2}
\end{figure}

PRISMA (Preferred Reporting Items for Systematic Reviews and Meta-Analyses)~\cite{turrini2017common} is a prominent and extensively used approach for presenting a systematic review~\cite{moher2015preferred}. This study reported on a PRISMA-based systematic literature review (SLR). Paper searching, screening, and selection were carried out in accordance with the PRISMA 2015 guidelines, as indicated in Figure 1. The first search was conducted using two major datasets: Scopus and Web of Science (WOS). Many scholars and practitioners rely on Scopus and WOS databases for locating quality articles and SLR~\cite{ahsan2021machine,mustapha2021impact,manjurul2021machine,ahsan2022machine}. To keep the study aim more focused and key oriented, we used the Boolean operator with "health care" and "Industry 4.0" during the initial search, yielding 869 articles from Scopus and 378 articles from WOS. When the article is restricted to journal papers, peer review, and English, the whole article after title, keyword, and abstract analysis reached to 346.\\
One researcher (Z.S.) imported data from 346 journal articles into Excel CSV files in preparation for a more in-depth investigation. Using Excel's duplication tools, duplicates were discovered and eliminated~\cite{manjurul2021machine}. Two independent reviewers evaluated the titles and abstracts of 346 articles (M.A. and Z.S.). To eliminate bias, disputes about article selection were settled by discussion and agreement. Based on the inclusion-exclusion criteria indicated in Fig.~\ref{fig:fig2}, a total of 32 papers were identified for qualitative synthesis.

\section{Bibliometric analysis}\label{analysis}
The following section describes the bibliometric analysis of selected 346 articles. The bibliometric analysis was carried out using Excel data analysis, R-Studio software, and VoSviewer software. The bibliometric analysis includes the analysis of the articles based on the article's publication by year, discipline, journals, citations, authors, countries, institutions, keywords, and co-authorship analysis.
\subsection{Publication by year}
\begin{figure}[htbp]
    \centering
    \includegraphics[width=\textwidth]{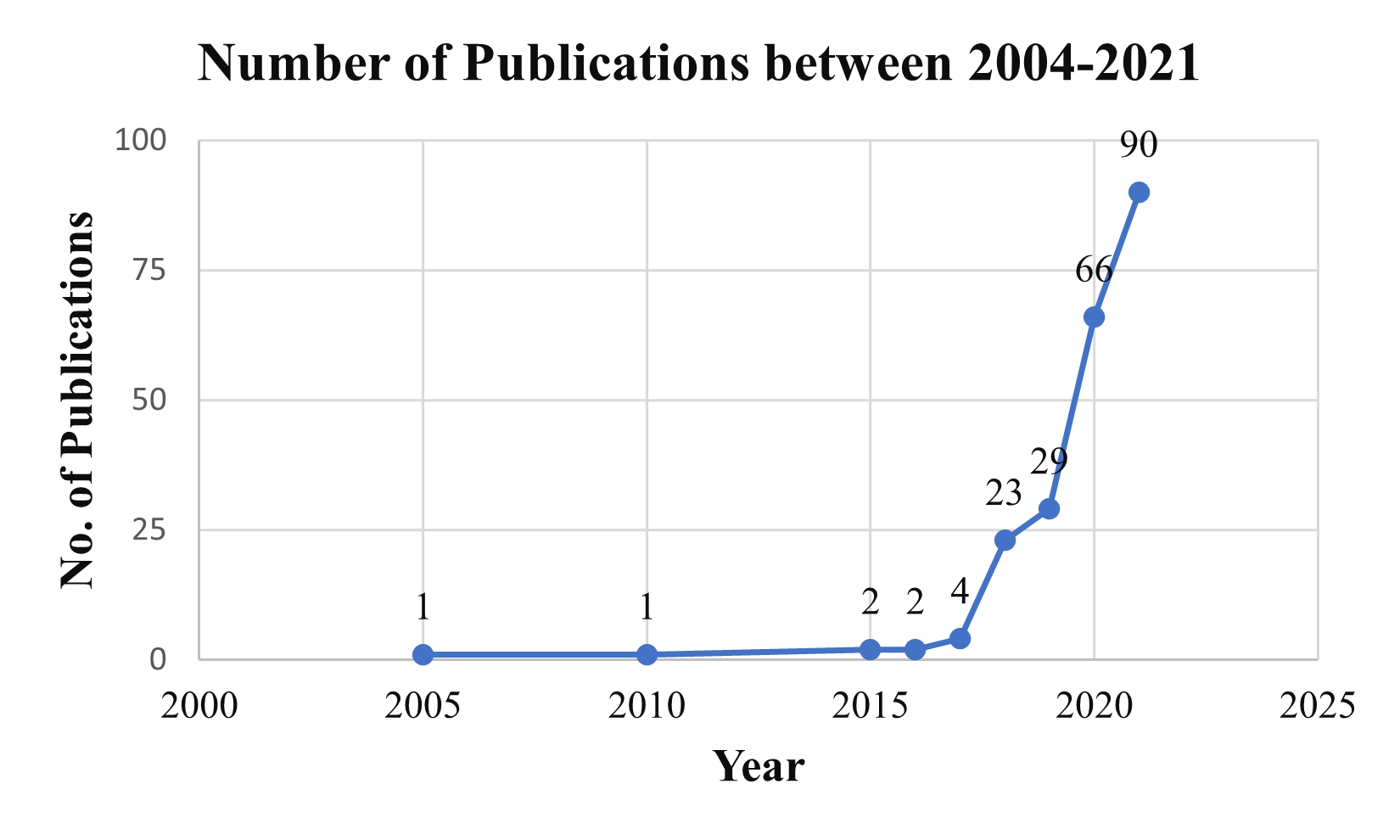}
    \caption{Publications by year}
    \label{fig:fig3}
\end{figure}

Fig.~\ref{fig:fig3} depicts the large increase in research article publications over the previous seven years, demonstrating a higher degree of interest in the academic community in Industry 4.0 in healthcare (IHC). The statistic shows that the published literature was nearly non-existent between 2005 to 2015. The number of published publications in 2021 is expected to be over 90, while in 2015 it was only 2. We anticipate that the exponential rise in IHC publication will continue in 2022 and later years.
\subsection{Publication by discipline}
\begin{figure}[htbp]
    \centering
    \includegraphics[width=\textwidth]{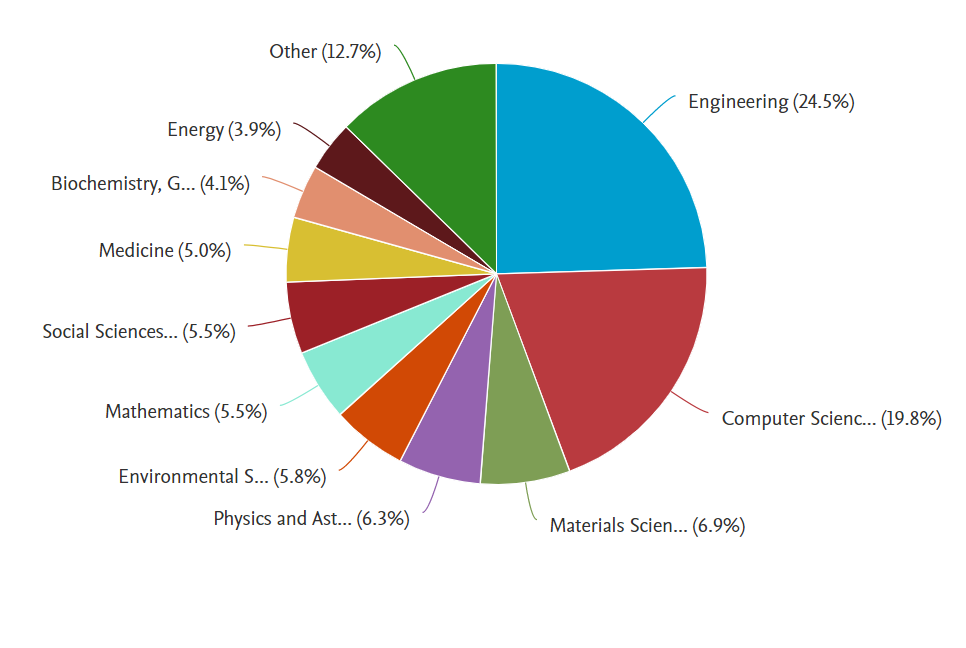}
    \caption{Documents by subject area}
    \label{fig:fig4}
\end{figure}

Fig.~\ref{fig:fig4} depicts the IHC literature in a range of subject areas. IHC research is dominated by engineering and computer science, which account for 24.5 percent and 19.8 percent of total publications, respectively. Furthermore, interest in IHC has increased in other disciplines such as material science, mathematics, and so on.
\subsection{Publication by journals}
\begin{figure}[htbp]
    \centering
    \includegraphics[width=\textwidth]{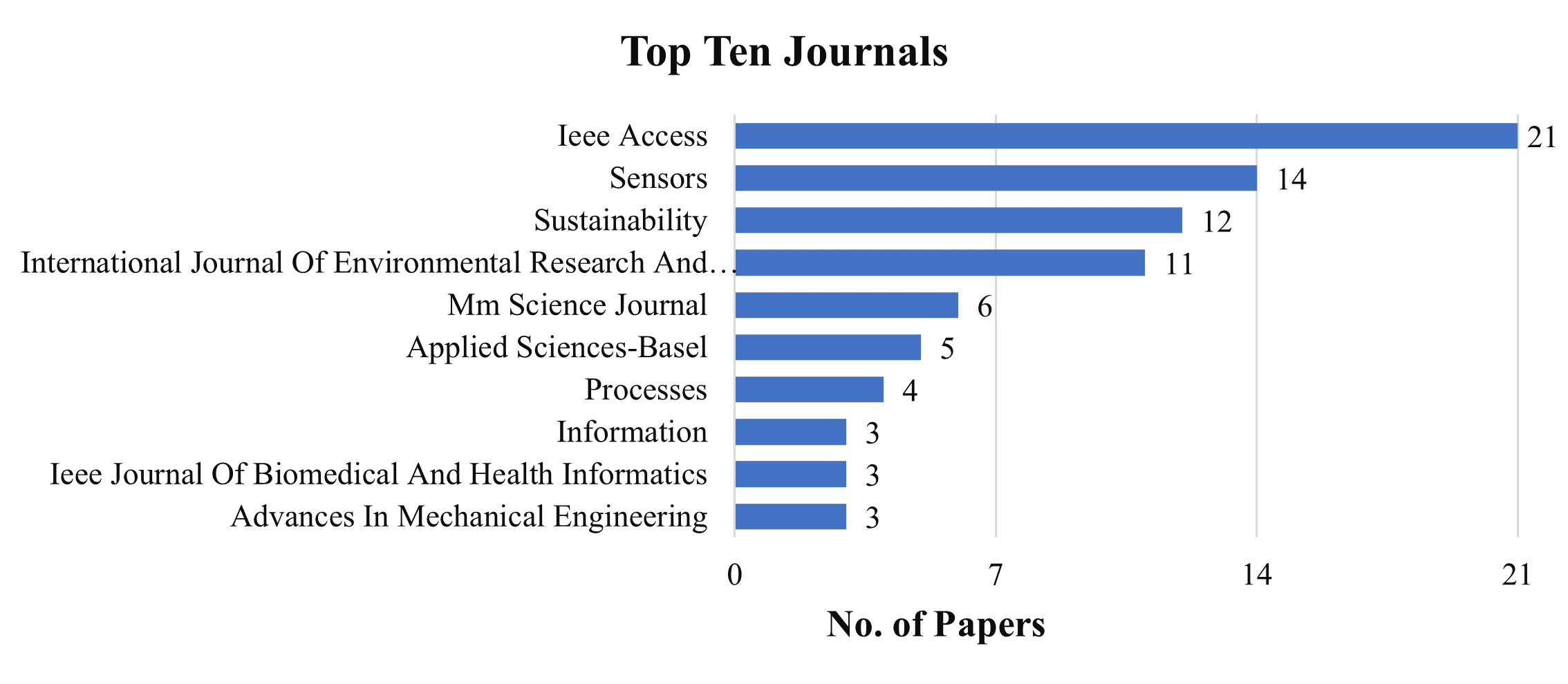}
    \caption{Publications by Journals}
    \label{fig:fig5}
\end{figure}

We examined the most productive articles in the IHC domains. Fig.~\ref{fig:fig5} illustrates the top ten journals and the overall number of articles published during the preceding decades. According to referenced literature, the three most productive journals are IEEE Access (21 publications), Sensors (14 articles), and Sustainability (12 papers). Surprisingly, just 13.5 percent of all cited content was published in the top three publications.
\subsection{Publications by citations}
The study evaluates the overall amount of citations while collecting data and thoughts on IHC's renowned authors.
Table~\ref{tab:tab1} displays the ten most frequently referenced publications according to the Scopus and WOS databases. Table~\ref{tab:tab1} shows that each of the authors obtained citations between 44 and 184. Because various databases utilize different indexing techniques and time periods, the total number of citations may differ from Google Scholar and other database citations. According to the Table, the paper published by Del et al. (2019) obtained the most citations (184), with 46 citations each year, followed by the article published by Mittal et al. (2019), which earned 154 citations in total (38.5 citations per year). All of the articles in Table 1 are likely to be among the most prominent articles in the IHC.
\begin{table}[htbp]
    \centering
     \caption{Top 10 cited papers published in MLBDD between 2012-2021}
    \begin{tabular}{p{.17\linewidth}p{.4\linewidth}p{.1\linewidth}p{.17\linewidth}}\toprule
       Author(s)&Article Titles&	Citation&	Citation/Year\\\midrule  
      Del et al. (2019)~\cite{del2019bio}&Bio-inspired computation: Where we stand and what's next&
	184&	46\\
Mittal et al. (2019)~\cite{mittal2019smart}& Smart manufacturing: Characteristics, technologies and enabling factors&
	154&	38.50\\
Pace et al. (2018)~\cite{pace2018edge}&An edge-based architecture to support efficient applications for healthcare Industry 4.0&
	149&	37.25\\
Javaid et al. (2020)~\cite{javaid2020industry}&Industry 4.0 technologies and their applications in fighting COVID-19 pandemic&
	137&	45.67\\
Reis \& Gins (2017)~\cite{reis2017industrial}&Industrial process monitoring in the Big Data/Industry 4.0 Era: from detection, to diagnosis, to prognosis&
	104&	17.33\\
Grapov et al. (2018)~\cite{grapov2018rise}&Rise of deep learning for genomic, proteomic, and metabolomic data integration in precision medicine
	&83&	16.6\\
Gupta et al. (2020)~\cite{gupta2020smart}&Smart contract privacy protection using AI in cyber-physical systems: tools, techniques and challenges
	&50&	16.67\\
Abdel et al. (2021)~\cite{abdel2021intelligent}&An intelligent framework using disruptive technologies for COVID-19 analysis
	&50&	25\\
Khalid et al. (2018)~\cite{khalid2018security}&Security framework for industrial collaborative robotic cyber-physical systems
	&46&	9.20\\
Pilloni et al. (2018)~\cite{pilloni2018data}&How data will transform industrial processes: crowdsensing, crowdsourcing and big data as pillars of industry 4.0&
	44&	8.8\\\bottomrule

    \end{tabular}
   
    \label{tab:tab1}
\end{table}
\subsection{Publication by authors}
Table~\ref{tab:tab2} lists the top ten authors who have published the most publications on IHC during the last few decades. Kumar Satish, in comparison to the other authors, published the most papers (4). Authors Khan, Li, Park, and Zalud placed second having three papers.
\begin{table}[htbp]
    \centering
    \caption{Most influential author in terms of total publications}
    \begin{tabular}{ccc}\toprule
         Authors&	No. of Articles& Articles Fractionalized \\\midrule
Kumar S&	4&	0.92\\
Khan M&	3&	0.67\\
Li Y&	3&	0.58\\
Park H&	3&	0.92\\
Zalud L	&3&	0.88\\
Ahmad M	&2&	0.46\\
Ajayi O	&	2&	0.67\\
Algarni F&	2&	0.42\\
Andonegui I&	2&	0.50\\
Bagula A&	2&	0.67\\

\bottomrule
    \end{tabular}
    
    \label{tab:tab2}
\end{table}
\subsection{Publication by countries}
Table~\ref{tab:tab3} summarizes the top ten most productive countries. We discovered that Italy, the United Kingdom, and China are the most influential nations in scientific production when examining the leading countries. According to Table~\ref{tab:tab3}, Italy has 57 published papers, followed by the United Kingdom (35), China (31) and so on.
\begin{table}[htbp]
    \centering
    \caption{Top ten most productive countries}
    \begin{tabular}{cc}\toprule
Country&	Articles\\\midrule
Italy&	57\\
UK&	35\\
China&	31\\
India&	29\\
Spain&	29\\
USA&	28\\
Germany&	24\\
Brazil	&23\\
Czech Republic&	19\\
Australia&	14
\\\bottomrule

    \end{tabular}
    
    \label{tab:tab3}
\end{table}
\subsection{Publication by institutions}
\begin{figure}[htbp]
    \centering
    \includegraphics[width=\textwidth]{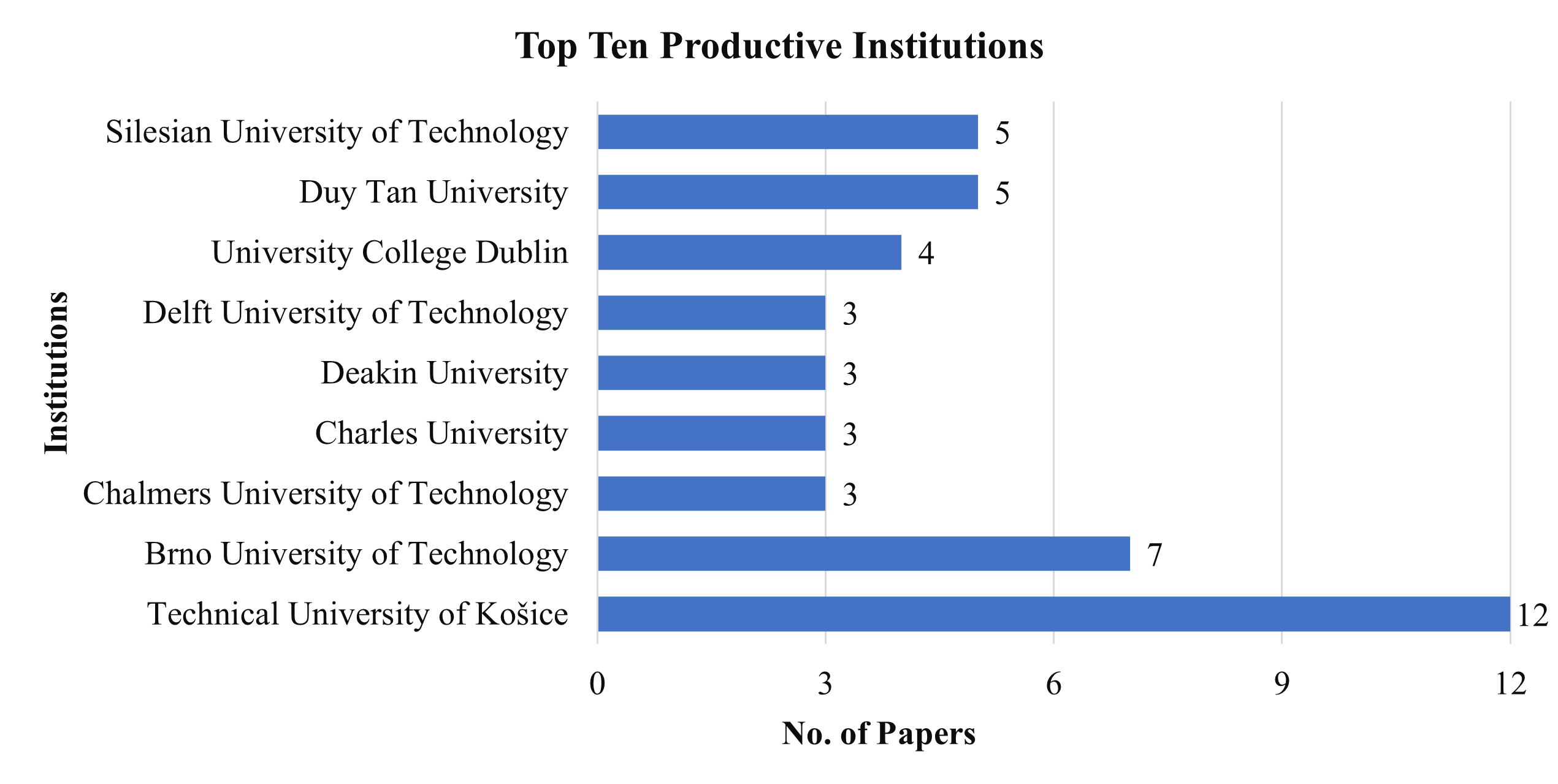}
    \caption{Top ten institutions based on number of publications}
    \label{fig:fig7}
\end{figure}

According to our investigation of the top ten academic institutions, the Technical University of Kosice (Slovakia) is the most productive, with 12 (approximately 3.46 percent) papers published, as shown in Fig.~\ref{fig:fig7}, followed by Brno University of Technology (Czechia) with 7 (2.02 percent) published articles.
\subsection{Common words in keywords}
\begin{figure}[htbp]
    \centering
    \includegraphics[width=\textwidth]{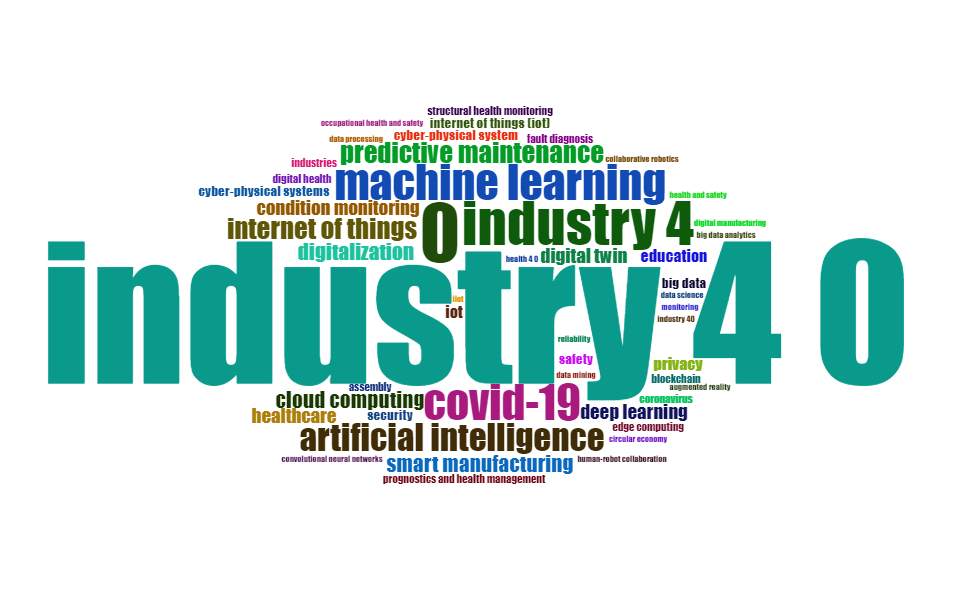}
    \caption{Word cloud for most frequently used keywords in MLBDD publications}
    \label{fig:fig9}
\end{figure}

\begin{figure}
    \centering
    \includegraphics[width=\textwidth]{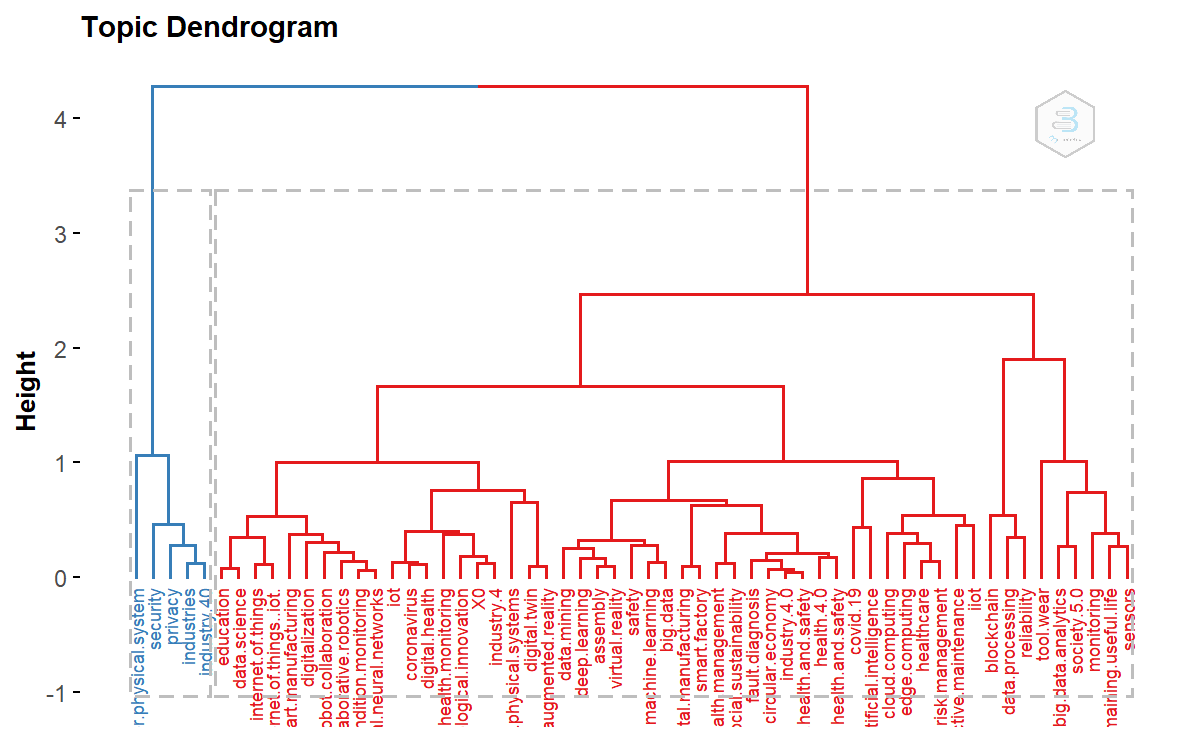}
    \caption{Dendrogram of Authors keywords}
    \label{fig:dendo}
\end{figure}

The most commonly used terms by researchers in the keywords section were also analyzed because they typically express the study's core topics. R-studio software was used to conduct the analysis. Fig.~\ref{fig:fig9} illustrates the most frequently used terms in the keyword sections of 346 referenced literature. "Industry 4.0," "COVID-19," "artificial intelligence," and "machine learning" are among the most often used phrases, as evidenced by their high and strong font visibility.
\\
The equivalent analysis and clustering dendrogram for keywords is shown in Fig.~\ref{fig:dendo}. In this case, the height indicates the space between the words and the space indicates how each concept differs from the others.
\subsection{Co-authorship of countries}
For the co-authorship of countries analysis, the minimum number of publications for a nation was set to 5, demonstrating the collaboration of authors from diverse geographical locations. As a result, 19 of the 63 countries satisfied the requirement. Figure 2 demonstrates that nations that are geographically close to one another have a strong affinity. Based on the co-authorship analysis, we discovered that the United States has the most connections with other countries, with 10 linkages and a total link strength (TLS) of 20, as well as 26 published articles and 2855 citations. India (10 connections, 18 TLS, 23 documents, and 875 citations) and China were next ( Links: 9, TLS: 17, documents: 24 and citations: 290). Table~\ref{tab:coauthor} contains a list of all 19 nations, as well as their relationship strength, number of documents, and total citations, while Figure 6 depicts a snapshot from the VOSviewer demonstrating how the countries are connected. Scientific collaboration is seen as a critical component in improving the quality and impact of research~\cite{ozbaug2019bibliometric}. Expanding international collaboration may be accomplished through a variety of methods, including increasing the number of visiting researchers, forming diverse collaborations, and investing substantial funds in research. To enhance international ties, a comprehensive and adaptable research policy is required~\cite{khudzari2018bibliometric}.
\begin{table}[htbp]
  \caption{Analyses of countries' Co-authorship}
    \centering
    \begin{tabular}{cccc}\toprule
	Country&	Documents&	Citations&	Total Link Strength\\\midrule
	United States&	26&	2855&	20\\
India&	23&	875&	18\\
China&	24&	290&	17\\
Australia&	12&	88&	17\\
Pakistan&	12&	217&	17\\
United Kingdom&	18&	591&	13\\
Saudi Arabia&	7&	281&	12\\
Italy&	31&	469&	9\\
Germany&	18&	249&	9\\
Brazil&	14&	132&	9\\
Spain&	14&	318&	7\\
Malaysia&	6&	6&	6\\
Portugal&	6&	187&	5\\
Canada&	9&	177&	4\\
South Korea&	7&	176&	4\\
France&	6&	102&	4\\
Poland&	6&	33&	3\\
Taiwan&	6&	49&	2\\
Indonesia&	5&	65&	2\\\bottomrule

    \end{tabular}
  
    \label{tab:coauthor}
\end{table}
\begin{figure}[htbp]
    \centering
    \includegraphics[width=\textwidth]{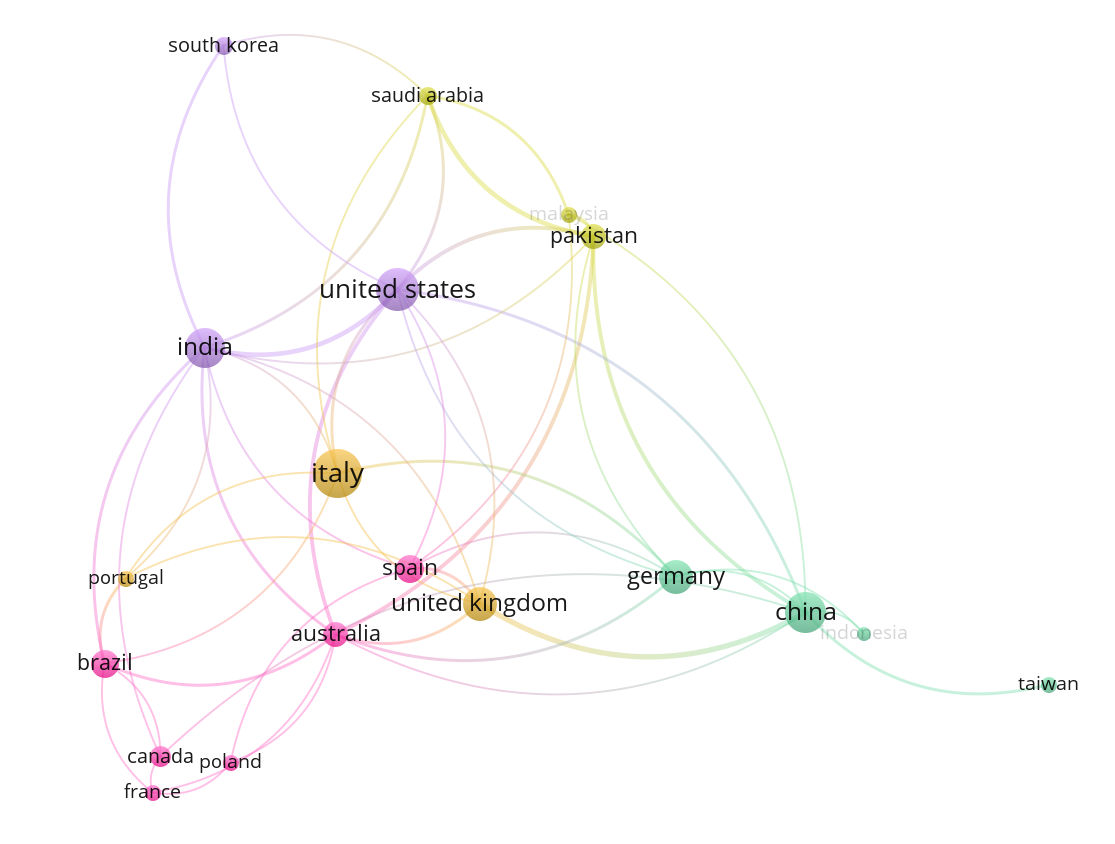}
    \caption{Co-authorship of countries}
    \label{fig:fig10}
\end{figure}
\subsection{Co-authorship (authors)}
\begin{figure}
    \centering
    \includegraphics[width=\textwidth]{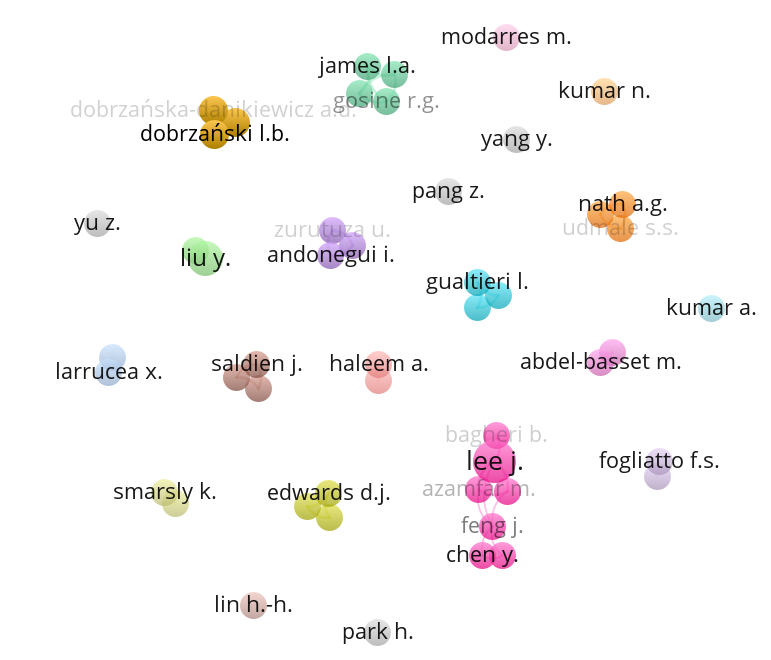}
    \caption{The bibliometric map depicting the analysis of authors' co-authorship}
    \label{fig:fig11}
\end{figure}

For the research of author co-authorship, a network of co-authors is discovered, and authors who collaborated are identified. Co-authorship analysis~\cite{chen2019international} is the most often used method for analyzing research collaboration (RC)~\cite{callon1983translations}. Researchers' total co-authorship ties are shown in the links. A researcher's co-authorship connection with other researchers~\cite{van2020vosviewer} is shown using total link strengths (TLS). The minimum number of documents and citations for one author was set to two and ten, respectively, for the purposes of author co-authorship analysis. Only 49 out of the 773 authors met the criteria. As a result of the close connections among the 49 authors, different distinct groups emerged (as displayed in Fig.~\ref{fig:fig11}). This group of six authors has a wide range of expertise and a long history of working together. Table~\ref{tab:coacoa} displays the links, TLS, documents, and citations for top ten authors. Lee J. has four published articles and 2478 citations and has the highest total link strength (TLS=7), followed by Gosine R. (TLS=6, documents=2, and citations=19), and James L. (TLS=6, documents=2, and citations=19).
\begin{table}
\caption{Co-authorship analysis of authors}
    \centering
    \begin{tabular}{cccc}\toprule
         Author&	Documents&	Citations&	Total link strength\\\midrule
Lee j.&	4&	2478&	7\\
Gosine R.&	2&	19&	6\\
James L.&	2&	19&	6\\
Wanasinghe T.&	2&	19&	6\\
Warrian P.&	2&	19&	6\\
Azamfar M.&	2&	91&	5\\
Singh J.&	2&	91&	5\\
Dobrzańska A.&	2&	13&	4\\
Dobrzański L.&	2&	13&	4\\
Dobrzański B.&	2&	13&	4\\\bottomrule

    \end{tabular}
    
    \label{tab:coacoa}
\end{table}
\subsection{Co-occurrence author keywords}
\begin{figure}
    \centering
    \includegraphics[width=\textwidth]{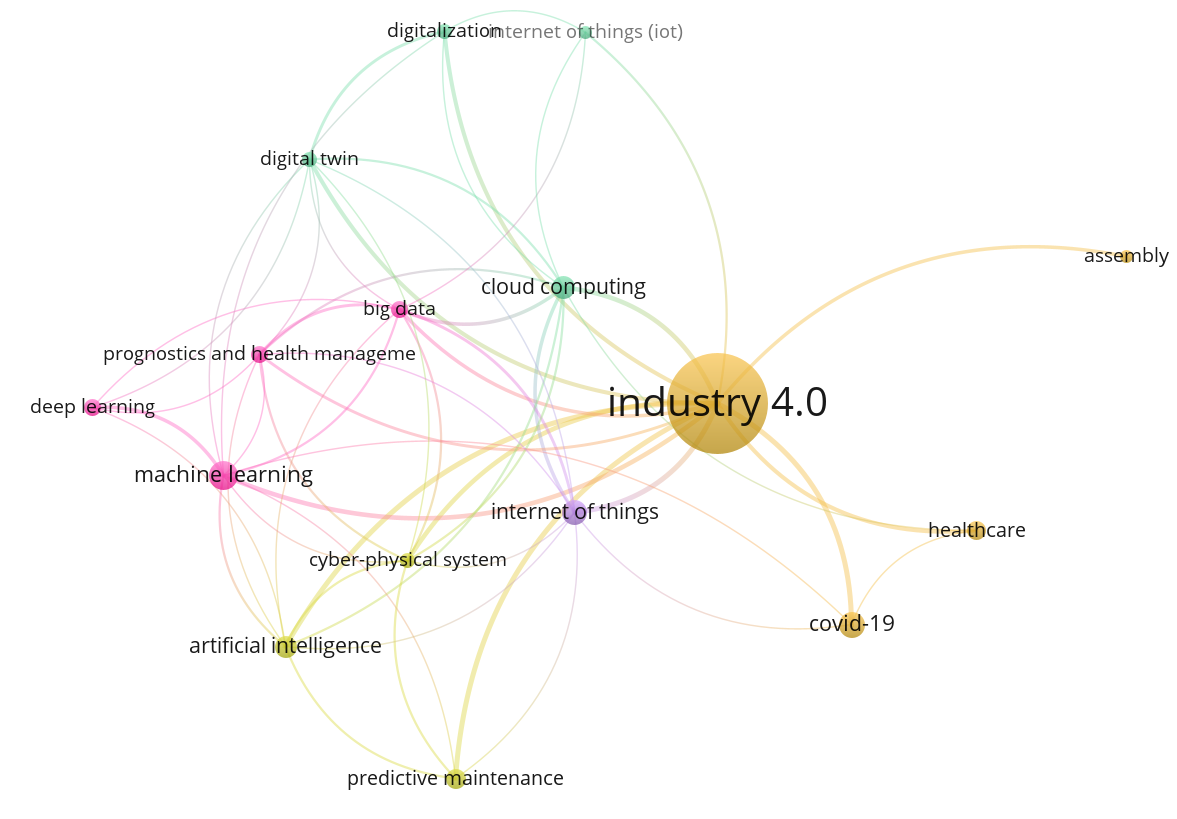}
    \caption{Co-occurence of author keywords networks developed by VOSviewer}
    \label{fig:cooca}
\end{figure}

 The author's keyword sections convey information in a unique way. "Co-occurrence" refers to the frequency with which words appear in specific publications~\cite{callon1983translations,khan2020systematic}. The total length of a word's strength indicates the number of times it appears in a document. The node size represents the frequency of the words; for example, the larger the node size, the more common the phrase. A thicker line connecting two or more words shows that they are close to a cluster. Scopus data were imported into the VOSviewer application for author keyword co-occurrence analysis. With a minimum of five keyword occurrences, 385 of 4230 keywords were found. Each of the 385 keywords was classified into one of five clusters. Vosviewer's co-occurrence networks are depicted in Fig.~\ref{fig:cooca}. For instance, Keywords in Cluster 1 include big data, deep learning, machine learning, prognostics, and health management.
\subsection{Bibliographic coupling (documents)}
\begin{figure}
    \centering
    \includegraphics[width=1.01\textwidth]{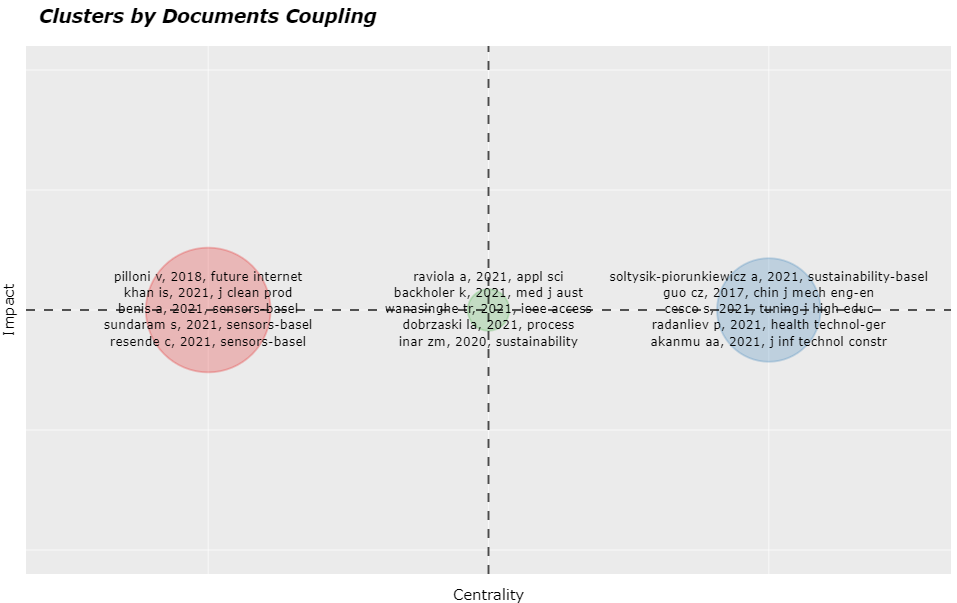}
    \caption{The bibliometric map representing documents coupling}
    \label{fig:fig12}
\end{figure}

The R-Studio application was used to conduct the bibliographical coupling analysis. The linking of studies that reference each other is an essential aspect of bibliographic coupling. Our unit of analysis had been determined, and our counting technique had been meticulously followed. The minimum number of units for this study is 25, the minimum clustering frequency is 5, and the number of labels per cluster is 5. The wider the circle in Fig.~\ref{fig:fig12}, the stronger the bibliographic connection appears to be. The most prominent cluster was discovered in articles addressing keywords such as future internet, clean manufacturing, and sensor, with a total frequency of 11 and a centrality of 3.91.
\subsection{Thematic evaluation}
\begin{figure}[htbp]
    \centering
    \includegraphics[width=1.05\textwidth]{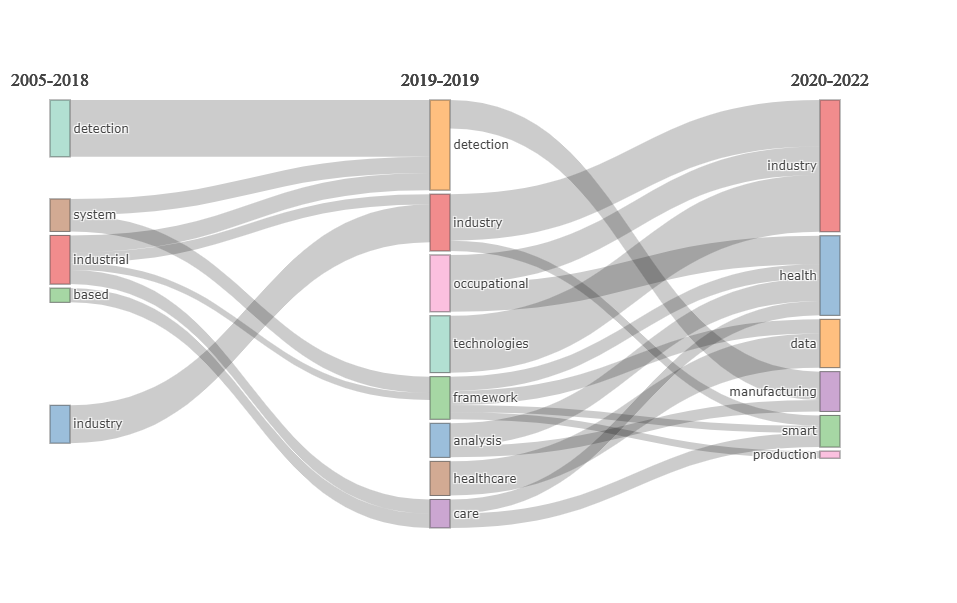}
    \caption{Thematic assessment of the authors' notion based on title analysis}
    \label{fig:fig13}
\end{figure}

Based on the distribution of publications each year, we divided our collection into three time periods: 2005-2018, 2019-2019, and 2020-2022. Fig.~\ref{fig:fig13} illustrates a thematic assessment of the authors' concept based on title analysis. Topics such as industry, system, and industrial began as specialized themes, and by 2019-2019, a different topic such as technology, healthcare, care, and the framework merge with those topics. Manufacturing, smart, and production become significant ideas in IHC between 2020 and 2022, as seen in Fig.~\ref{fig:fig13}.
\subsection{Historiographic mapping}
\begin{figure}[htbp]
    \centering
    \includegraphics[width=\textwidth]{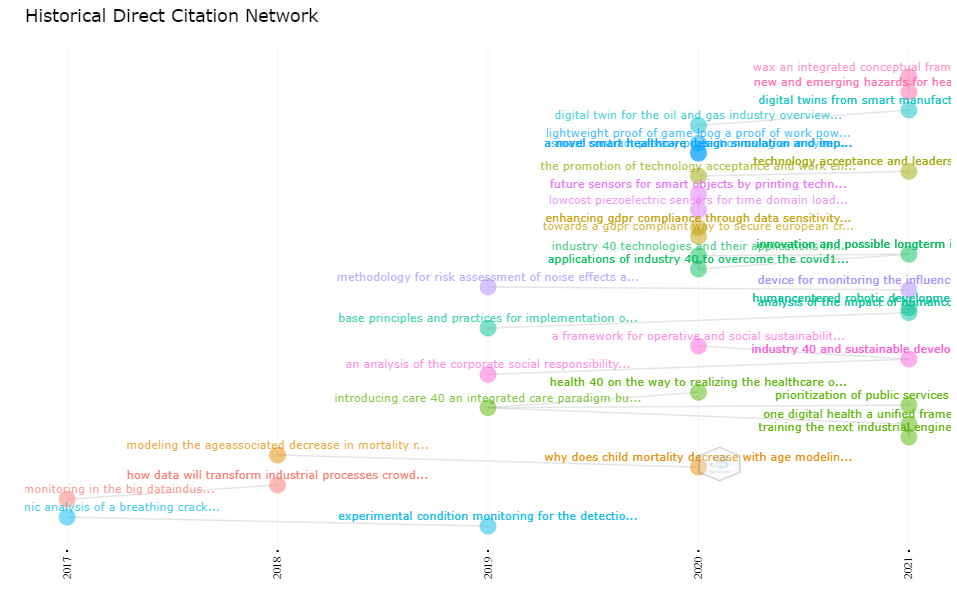}
    \caption{Historiographic mapping}
    \label{fig:fig14}
\end{figure}

Fig.~\ref{fig:fig14} depicts a historiographic mapping based on the document titles. Each node in the graph indicates a document from the examined dataset that has been mentioned by another document. In addition, each title is represented by a distinct edge. For example, Chute \& French (2019),  published a paper in 2019 that was cited in one publication in 2020 and three articles in 2021. 
\section{Insights of IHC}\label{HCI}
\begin{figure}[htbp]
    \centering
    \includegraphics[width=\textwidth]{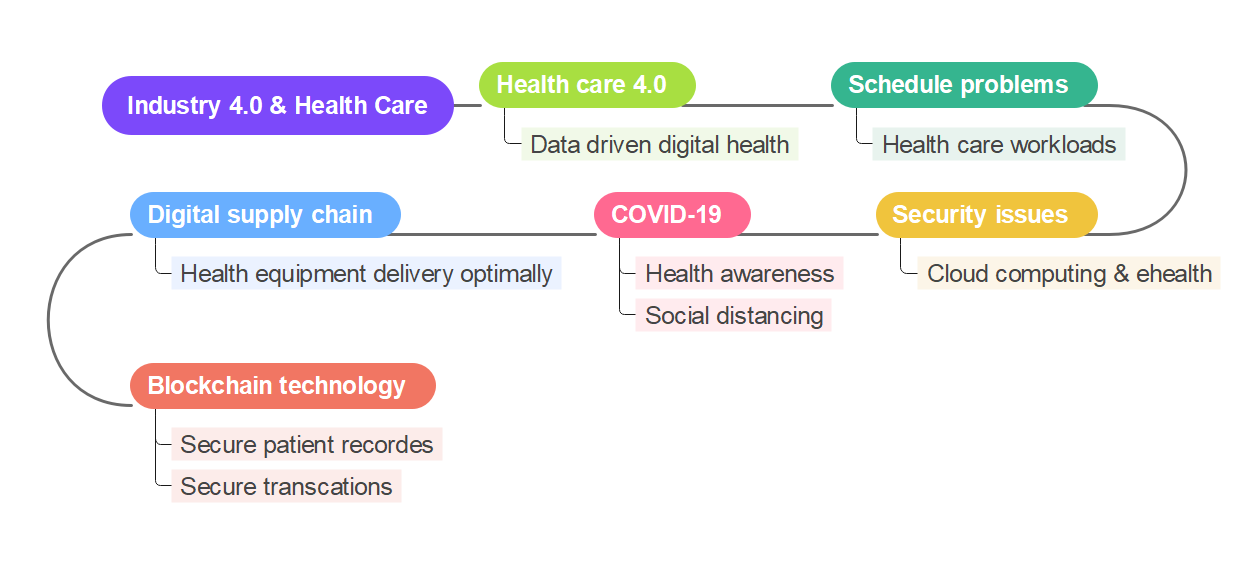}
    \caption{Outcomes of IHC based on referenced literature}
    \label{fig:frame}
\end{figure}

In this section, industry 4.0 and health care (IHC) have been analyzed in detail based on the selected 32 articles. The following section presents the papers' classification according to six major categories in IHC. The literature has been classified as (i) conceptual framework health care 4.0, (ii) schedule problems, (iii) security issues, (iv) COVID-19, (v) digital supply chain, and (vi) blockchain technology. Fig.~\ref{fig:frame} summarizes the overall classification of IHC based on the referenced literature.
\subsection{Conceptual framework health care 4.0}
Health care has been around since the beginning of civilization, but in the last few decades, it has evolved dramatically. Health care 1.0 was largely limited to providing physical health services, such as checking vital signs and delivering babies via natural means. Then, with the advent of modern medicine, health care 2.0 added mechanical services like surgery and medical machines to its repertoire. Health care 3.0 patient-centric, and health care 4.0 deals with cloud computing, fog computing, and IoT. With growing population and technological facilities, it is becoming pertinent to facilitate the health care service with better facility and flexible manner. According to Jayaraman et al. (2019) Healthcare 4.0, is a relatively new phrase developed from Industry 4.0~\cite{jayaraman2020healthcare}. In healthcare 4.0, data-driven digital health technologies include innovative health, online health, mHealth, mobile health, wifi health, eHealth, telehealth/telemedicine, medical information technology digital medicine, health informatics, ubiquitous health, and health information systems~\cite{jayaraman2020healthcare}.\\
The emerging COVID-19 demonstrated why it is critical to consider IHC when developing various approaches that may serve both patients and practitioners while providing security, accountability, and system efficiency—digital technology aids in the establishment of a safe working environment for healthcare professionals~\cite{kaiser2021iworksafe}. A few studies that have been mentioned provide a conceptual underpinning for more effective and efficient health-care systems. Dobi et al. (2019), for example, provide a novel technique that is appropriate for use in the health care business, specifically for monitoring a patient's health features. The framework is built using Markov chain-based techniques. The Markov chain's stationary distribution may be used to estimate the average cost of the procedure, which is determined by the time between samplings and the control limit.\\
Benis et al. (2021) presented One Digital Health, a cohesive framework for long term health environments . The conceptual framework for One Digital Health is based on two core concepts (One Health and digital Health), three perspectives (individual health and well-being, people and community, and environment), and independent variables (citizen engagement, education, environment, human and veterinary health care, and Healthcare Industry 4.0). One Digital Health's goal is to digitalize long-term health settings through the adoption of a systematic life and health sciences approach that considers multiple digital technology views on public health, animal health, and environmental management. Other issues, such as the exchange of electronic health information, are, however, unresolved within the jurisdiction of primary health care systems~\cite{larrucea2020towards}.
\subsection{Schedule problems}
Scheduling problems entail determining the ideal schedule given various objectives, system conditions, and job characteristics. While manufacturing plant, garments industry extensively faces schedule problems, during the last decades, health care systems have also considered schedule problems one of the key factors that cause the bottleneck of the smart health care systems. Several studies identified schedule problems in health care sectors and proposed a new scheme to solve those challenges. Dootio et al. (2022) developed the lightweight remote procedure call (RPC) for the IIoHT mechanism. The study dealt with schedule problems and scheduled all healthcare workloads under their deadlines. The study uses linear integer programming that helps to reduce the schedule delay up to 50\% compared to the existing RPC~\cite{dootio2021secure}.\\
Chauhan et al. (2021) established and studied seven principles linked to smart healthcare waste disposal systems with circular economy characteristics to retrieve value of disposables using a decision making trial and evaluation laboratory (DEMATEL) technique. A causal diagram was used to select the criteria based on their relevance and net cause and effect connection~\cite{chauhan2021interplay}. \\
Most of the smart health services are hospital-oriented. Additionally, patients' data are not organized and application-dependent, causing attention delays in many cases. To overcome such scenarios, Bedon et al. (2020) developed a home-based smart health model considering the Internet of things (IoT), Internet of medical things, and smart health applications. The proposed health service reduced the attention delays compared to existing health care facilities~\cite{bedon2020home}.\\
Ajayi et al. (2019) aim to solve the scheduling problem in Africa by developing interconnected medical facilities across the nations. A cooperative and competitive collaboration model was developed considering two allocation schemes such Genetic Algorithm-based VM Allocation (GAVA) and Stable Roommate Allocation (SRA)~\cite{ajayi2020africa}. However, one of the limitations of that study is that the study result was based on the simulation model. Therefore, the question was raised regarding the performance of the proposed model in real-world scenarios.  
\subsection{Security issues}
Patient data privacy is one of the most urgent issues in the health-care sector. According to a 2020 study published in Cybe-rcrime magazine, 93 percent of U.S. healthcare organizations have experienced a severe security breach that resulted in data breaches, denial-of-service attacks, malware, or other cyber-intrusions. It is impossible to determine how frequently individuals' EHR (electronic health record) systems have been hacked without doing an audit of their EHR (electronic health record) system~\cite{cyber2020}. Cloud settings have become one of the most enticing areas for hackers since the bulk of health care systems store patient data on the cloud. In view of the security difficulties in IHC, multiple studies suggested several protective techniques to secure cloud-based medical data.\\
As a result, the health-care business is working to develop a cloud server that requires a higher level of security and authenticity. Signcryption, a mix of signature and encryption, for example, has demonstrated promising outcomes in cloud-based access systems. The bulk of them, however, are better suited to homogeneous rather than heterogeneous situations. Ullah et al. (2022) developed an access control mechanism for IoT settings that leverages heterogeneous signcryption to overcome these issues. According to the authors, their proposed technique might survive a range of risks such as secrecy, integrity, and so on~\cite{ullahaccess}.
Khan et al. carried out a similar research (2022). Furthermore, the author created an efficient signcryption technique for IoT. The Internet of Things makes use of the Internet as a key channel of communication for single documents as well as multi-digital communications. The author investigated the following security characteristics in terms of processing cost and transmission bandwidth: anonymity, untraceability, confidentiality, and unforgeability. The author developed a lightweight and secure proxy blind signcryption for multi-digital communications using a hyperelliptic curve (HEC), which is less computationally costly and provides higher transmission capacity~\cite{khan2022efficient}.\\

Care 4.0, an integrated care paradigm presented by Chute et al. (2019), aims to provide patients the maximum control over their medical information. Care 4.0's overarching objective is to build trustworthy and connected networks that are linked with digital health and care services. The planned care 4.0 networks and technology may assist individuals in managing and utilizing their assets, both inside their local care circle and throughout the larger community~\cite{chute2019introducing}. In addition to Chute et al., Pace et al. (2019) proposed an efficient application for healthcare industry 4.0. A simple client module that supports multi-radio and multi-technology is constructed to collect local data. One of the potential advantages of their proposed architecture is that the capabilities are accessible via both private and public cloud platforms, providing for better flexibility and resilience. In real-time testing, the proposed design beat competing devices in terms of data transmission and processing time~\cite{pace2018edge}. \\
Rather than that, Alvarado et al. (2020) defines Healthcare 4.0 as the adoption of three key paradigms: the Internet of Things, Big Data, and Cloud Computing, all of which are changing eHealth and its whole ecosystem in the same manner as Industry 4.0 is modernizing manufacturing~\cite{borregan2020bibliometric}.
\subsection{COVID-19}
Over 800 million people have been left without basic health care as a result of the present COVID-19 crisis, which has resulted in a worldwide epidemic of life-threatening illnesses. While the death toll is believed to exceed 55,000,00, with an additional 2 million hospitalized each week, this problem is not going away anytime soon~\cite{worldmeter,ahsan2020covid,ahsan2020deep}. Mobile applications, robots, Wi-Fi cameras, scanners, and drones are being utilized to combat virus propagation. Industry 4.0 has had an impact on the important contribution of digital technology to pandemic control~\cite{narayanamurthy2021impact}.\\
The world was in the grip of an unprecedented health-care crisis at the height of COVID-19. As a result, academics and practitioners develop a wide range of tools and techniques to digitalizing systems in order to minimize disease transmission and increase the number of health care facilities. Majeed (2021), for example, developed a person-specific data collection strategy to decrease COVID-19 dissemination. One of their research's potential contributions is to use high-computing technologies to provide people with COVID-19-related knowledge in order to promote awareness. The recommended technique takes security into account when constructing prototype systems for users~\cite{majeed2021applications}.\\
Abdel et al. (2021) created an intelligent framework for COVID-19 analysis using disruptive technologies. The suggested approach, according to the authors, will help in the prevention of COVID-19 outbreaks, preserve the safety of healthcare staff, and protect patients' physical and psychological well-being. In response to the medical team's lack of personal protective equipment (PPE), the framework was created~\cite{abdel2021intelligent}.\\
According to Sust et al. (2020), disruptive technologies lessen the load on health personnel during the COVID-19 epidemic~\cite{sust2020turning}. Individuals infected with COVID-19 can be monitored utilizing cutting-edge monitoring technologies. In the instance of COVID-19, public health advocates argue that disruptive technology is essential for public health~\cite{benjamin2020ensuring}.\\
However, as the world is struggling to deal with COVID-19, it is assumed that the existing IHC systems are not enough to handle the situation. As of January 2022, the current COVID-19 infected cases is 307,685,974~\cite{worldmeter}, ultimately developing an extended crisis in health care. Thus, researchers and scholars should develop new solutions that might be more impactful in the current scenario in the health care sector.
\subsection{Digital supply chain}
The digital supply chain enables the timely and cost-effective delivery of the right product to the right patient~\cite{al2004extending}. The digital supply chain adds another layer to healthcare systems by lowering costs, reducing needless deviations, and increasing patient care and involvement. Dau et al. (2019), for example, offer a circular economy transition paradigm to investigate the healthcare sustainable supply chain 4.0. The proposed paradigm for health-care systems emphasized social responsibility~\cite{dau2019healthcare}.\\
During the SARS-COV-2 outbreak, Zahedi et al. (2021) established an IoT-based supply chain network. The proposed approaches shorten the response time of the ambulance. In addition, the proposed model minimizes the overall crucial reaction time. In Iran, the proposed supply chain model was tested in real-world scenarios. Over three weeks, the experimental findings show that the recommended methods reduced the COVID-19 case by up to 35.54 percent~\cite{zahedi2021utilizing}.\\
Fathollahi-fard et al. (2019) develop a bi-objective location-allocation-routing model for a Green Home Health Care Supply Chain (GHHCSC). The authors introduced five newly adopted annealing techniques using simulation approaches. Their final result demonstrates satisfactory results regarding the total cost of the system and environmental pollutions. However, as the proposed approach may not function as expected in recent times due to the ongoing crisis of COVID-19, yet required to modify the proposed scheme considering additional barriers (i.e., lockdown, transportation delay, route schedule)~\cite{fathollahi2019green}.\\
In addition, supply chain-related issues such as inaccurate inventory data, delayed supply allocations, slow repurposing of manufacturing facilities, and lack of innovations are ongoing challenges that need to be addressed in health care systems to develop dynamics IHC~\cite{sidhu2015digital,wickramasinghe2014lean}.
\subsection{Blockchain technology}
Blockchain has various applications in health care. For example, blockchain technology, has the potential to merge patients' medical information and prescription data from several locations into a single set of updated records~\cite{zhang2018blockchain,engelhardt2017hitching}. For example, Rupa et al. (2021) developed an Industry 5.0-based blockchain application to keep medical credentials on the Remix Ethereum blockchain~\cite{rupa2021industry}. Liu et al. (2017) suggested a new blockchain-based approach for a dependable and efficient medical record exchange system. One of the possible benefits of the suggested model is that it is a cost-effective technique that allows the user to immediately access the information without compromising security. However, o ne potential disadvantage is that the research does not address how existing e-health records would be integrated with the proposed architecture~\cite{liu2017advanced}.\\
Christo et al. (2019) advocated blockchain technology as a distributed solution to providing security while accessing medical reports. The proposed scheme contains authentication, encryption, and data retrieval. The authors claimed that the proposed framework protects the patients' records compared to other existing techniques~\cite{christo2019efficient}.\\
Kumar et al. (2020) demonstrated a wearable kidney system based on healthcare 4.0 protocols and utilizing blockchain technology. The author used game theory to design resource-constrained renal systems. The proposed system, according to the authors, would be an example of the shift away from specialist or departmental-centric methods and toward data and patient-centric approaches, bringing more transparency, trust, and good practices to the healthcare business~\cite{kumar2020lightweight}.
\section{Conclusions and remarks}\label{con}
Healthcare professionals have traditionally relied on technology to assist them give the best care to their patients, but significant advances in the previous decade have improved treatment quality and patient outcomes. Health care is always changing, but one constant remains: the necessity for a secure health care system free of data breaches. Today, health data may be found in electronic medical records (EMR), personal health records/digital health records, clinical data repositories (CDR), and health information exchanges (HIE). These archives hold essential information that must be protected, yet hackers routinely target it.
Because hackers are starting to target hospitals, healthcare organizations are finding that ensuring their IT infrastructure is intact and working is no longer enough. The fourth industrial revolution (IR) is the next generation of intelligent healthcare technologies that enable physicians and other professionals to more effectively manage disease transmission, protect their patients from dangerous illnesses, and receive real-time assistance when they are most needed. This study included a bibliometric examination of 346 publications on Industry 4.0 and health care (IHC). According to the conclusions of the survey, Italy is the most prominent country in terms of using Industry 4.0 in health care, followed by the United Kingdom and China. The three most productive journals in IHC in terms of total number of articles are IEEE Access, Sensors, and Sustainability. Kumar Satish has been regarded as the most productive author, with Khan, Li, Park, and Zalud following closely behind. The Technical University of Kosice is the most prolific institute, followed by the Brno University of Technology and the Chalmers University of Technology, which are placed second and third, respectively.\\
Finally, 32 papers were extensively reviewed to provide insights on IHC. IR 4.0 applications include scheduling difficulties, security, COVID-19, digital supply chain, and blockchain technology. However, some unresolved issues, such as reducing the carbon footprint, improving health care for aging populations, and securing long-term supplies during the pandemic, need prompt action. Data security, identity, and authenticity are still significant concerns in IHC, according to the referenced research. Researchers and practitioners may consider the limits identified in this paper as the area of future research. Future work might include, but is not limited to, explainable machine learning-based patient records, procedures and ethics for sharing patient data in cloud systems, and efficient machine learning and deep learning-based resource allocation in health care systems.
\bibliographystyle{unsrt}  
\bibliography{main}

\end{document}